\documentclass{article}
\usepackage{CJKutf8}
\usepackage[whole]{bxcjkjatype} 
\usepackage[unicode]{hyperref}
\usepackage{htnklab_bachelor}
\usepackage{graphicx}
\usepackage{color}

\usepackage{empheq}
\usepackage{amssymb}
\usepackage{amsmath}
\usepackage{algpseudocode}
\usepackage{algorithm}

\usepackage{float}
\english
\usepackage{color}

\usepackage{comment}

\usepackage{caption}

\newcommand{\argmin}{\mathop{\rm argmin}\limits}

\newcommand\scalemath[2]{\scalebox{#1}{\mbox{\ensuremath{\displaystyle #2}}}}
\DeclareMathOperator*{\argmax}{argmax} 

\etitle{Automatic Generation of Ice Hockey Defensive Motion via Coverage Control and Control Barrier Functions }

\eauthor{Kornvik Tanpipat, Takeshi Hatanaka, and Masatoshi Hiroura}

\englishabstract{
A successful defensive strategy in ice hockey games is often designed empirically by an experienced professional. The majority of previous work on automating the strategy focuses on analyzing spatial data to decide the most optimal formation and action but cannot generalize the system to real games with real-time capabilities. We propose a novel control logic for generating real-time ice hockey defensive motion based on a control barrier function (CBF) and coverage control to extend our antecessors' logic that succeeds in duplicating ideal formations for specific scenes. To this end, we first present an ellipsoidal CBF to overcome the drawbacks of the existing line-based CBF of our antecessors. We also tune and add a novel density function to reflect real specifications more precisely than the previous work. The control logic is then demonstrated through simulations with offensive motion in real games. It is confirmed that the present logic generates valid defensive movements without specification to these specific scenes. We further exemplify that the logic generates proper motion under ice hockey's man-to-man and zone defense strategies and their intermediate strategies by tuning the logic. This would contribute to reducing the efforts of the practitioners to educate ice hockey players.
}


\begin{document}

\maketitle

\section{INTRODUCTION}
Ice hockey is a sport in which players score goals in a six-on-six game on a field. Each team tries to achieve a goal by sending a puck, a black disk used in ice hockey, to the opponent's goal. Currently, the defensive formation against the opposing team's attack is determined empirically, and it is unclear whether it is indeed the optimal formation or not. Therefore, we seek to automate the creation of the defensive formation, utilizing the fact that the goal of the ice hockey defensive formation is to cover a large area while placing some defenders around the attacking players. Such action is similar to the objective of the coverage control problem, in which multiple robots form a network and move simultaneously while cooperating. Thus, we attempt to automate the simulation of ice hockey formations based on coverage control.

Our predecessors have approached the problem by proposing the above idea and succeeded in imitating defensive strategies for empirically designed specific scenes \cite{DAN2020}. They also simulate pass cut and goal block movements in these scenes by utilizing line-based control barrier function (CBF), which is a tool used to achieve a control's objective subject to safety guarantees \cite{CBFA2014}, \cite{CBFA2017}. However, their approaches are specific to some scenes. Moreover, Their line-based CBF is also not fully able to imitate pass cuts and goal blocks movements.

In this paper, we go beyond imitation to specific scenes. We propose a system to generate defensive formation for ice hockey in general scenes. We overcome the drawbacks of the previous logic regarding line-based CBF with a novel CBF model and also further improve the logic based on \cite{MDIC2021} and \cite{OOTA2020}. We also confirm that the novel system enables the generation of strategies in general scenes by inputting data from real scenes.

\section{PRELIMINARY}
In the following, we introduce the fundamentals of coverage control and control barrier functions for our simulation. These explanations are the same as mentioned in \cite{MDIC2021}.
\subsection{Coverage Control}
The coverage control is a multi-agent system control method that governs agents to move toward the optimal placement based on a predetermined density function. Many applications of coverage control have been reported in many fields, such as [4].
\label{cover}
\subsubsection{System model}
We consider the movement of n agents or defensive players $V \equiv \{1,\cdots,n\}$ on the 2-dimensional convex polyhedron field $Q \subset {\mathbb R}^2$. We also consider the defensive player model as the equation below
\begin{equation}
\label{dynamics}
 \dot{x}_i (t) = u_i (t),\, x_i(0) = x_{0i}
 \end{equation}
where ${x}_i = [x_{xi},x_{yi}]^T$ is the position of the $i^{th}$ defensive player and ${u}_i = [u_{xi},u_{yi}]^T$ is the input of the $i^{th}$ player. Each area in the field also has its weight, defined by the weight function $\phi(q)$.

\begin{figure}[htbp]

\centering

\includegraphics[width=0.6\linewidth,clip]{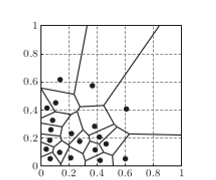}
\caption{Voronoi region by \cite{YHYF2018}}
\label{voronoi_fig}
\end{figure}

Next, we divide the field area $Q$ by the defensive players $x_i$ into several Voronoi cells $C_i$ according to Eq. \ref{voronoi}.
\begin{equation}
\label{voronoi}
 C_i(x) = \{q \in Q : ||q-x_i||^2 \leq ||q-x_j||^2, \forall{j} \in V  \}
 \end{equation}
 This implies that for any area $q \in C_{i}(x)$, the area is subjected to the cell, corresponding to the closest defensive player. For example, the visualization of the Voronoi cells can be seen Fig.\ref{voronoi_fig}.
 
\subsubsection{Evaluation function and control input}
 The evaluation function for the coverage control is defined as
\begin{equation}
\label{eval_func}
    J(x,t) = \sum_{i=1}^{n}{\int_{V_i(x)} ||q-x_i||^2\phi(q) \,dx}
 \end{equation}
 where $\phi(q)$ is the weight function of position $q$. This evaluation equation quantifies how ``well'' the agents cover the area given the predetermined weight function.

\begin{figure}[ht]

\centering

\includegraphics[width=0.6\linewidth,clip]{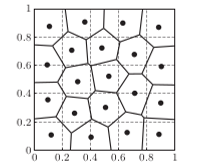}
\caption{Eventual result of coverage control by \cite{YHYF2018}}
\label{voronoi_eventual}
\end{figure}

With Eq. \ref{eval_func}, we calculate the equilibrium state of the agents corresponding to the evaluation function by solving its derivatives as shown below.
\begin{equation}
\label{eval_func_diff}
    \frac{\partial{J}}{\partial{x}}\Bigr|_{x=x^{*}} = 0
\end{equation}
By solving the above equation, we can derive the equilibrium state which is 
\begin{equation}
\label{eval_func_diff_ans}
    x_{i}^{*} = \frac{\int_{V_{i}(x)}q\phi(q)dq}{\int_{V_{i}(x)}\phi(q)dq}
\end{equation}
By considering the Voronoi cell $C_i(x)$ as a rigid body, we use Eq.\ref{mass_voronoi},\ref{cent_voronoi} to describe its properties.
\begin{eqnarray}
\label{mass_voronoi}
\rm{mass}(C_i(x(t))) &=& \int_{C_i(x)} \phi(q) dq  \\
\label{cent_voronoi}
\rm{cent}(C_i(x(t))) &=& \frac{1}{\mbox{mass}(C_i(x(t)))} \int_{C_i(x)} q\phi(q) dq 
 \end{eqnarray}
We can consider Eq. \ref{mass_voronoi} as the cell's mass representation and Eq. \ref{cent_voronoi} as the cells center of mass representation. Consequently, the interpretation shows that we can rewrite and consider the equilibrium state of the agents as the center of mass of each Voronoi cells which is 
\begin{equation}
\label{eval_func_diff_ans_cent} 
    x_{i}^{*} = \rm{cent}(C_i(x(t))) \nonumber
\end{equation}
As the objective of coverage control is to optimize the coverage of the agents over the area, we can design a controller that drives the agents toward the equilibrium state as Eq.\ref{control_input}.
\begin{equation}
\label{control_input} 
    u_i(t) = -k(x_i(t)-\rm{cent}(C_i(x(t))))
\end{equation}
In Eq. \ref{control_input}, $k$ is a constant gain that the designer can set. Thus, with the above controller, we can optimize the coverage of the agents over the specified area. For example, from Fig.\ref{voronoi_fig} and given that the weight function is some constant number over all the area., the eventual result of this setup is as shown in Fig.\ref{voronoi_eventual}. 

In summary, this section explains how the controller input is interpreted and calculated. The input will drive each agent toward its own Voronoi cell's centre of mass. For further details and proof regarding coverage control, please see \cite{YHYF2018}.

\subsection{Control Barrier Function (CBF)}
As stated in \cite{MDIC2021}, coverage control is necessary for the system to simulate goal block and pass cut movements. In this section, we introduce the fundamentals of CBF used in our simulation.

Considering the previously stated system dynamics from Eq.\ref{dynamics}
\begin{equation}
\label{new_dynamics} 
    \dot{x} = f(x)
\end{equation}
given $x \in D \subseteq \mathbb{R}^m $ as the system state and $f(x)$ as continuous Lipschitz function. Next, suppose we have a set $C$ that satisfy the following equations
\begin{eqnarray}
\label{on_in_bound}
C &=& \{ x \in \mathbb{R}^n :h(x) \geq 0 \},  \\
\label{on_bound}
\partial{C} &=& \{ x \in \mathbb{R}^n :h(x) = 0 \}, \\
\label{in_bound}
Int(C) &=& \{ x \in \mathbb{R}^n :h(x) > 0 \}
 \end{eqnarray}
given that $h : \mathbb{R}^m \rightarrow \mathbb{R}$ is a continuously differentiable function, we recall the following theorems due to \cite{MDIC2021}. 
\newline\newline
\textbf{Theorem 1:} \emph{From the system in Eq. \ref{new_dynamics} and set $C$ in Eq. \ref{on_in_bound}-\ref{in_bound}, if there exists class $K$ function $\alpha$, which satisfies the equation below, then $h$ function is a control barrier function over $C \subseteq D \in \mathbb{R}^m $}
\begin{equation}
\label{cbf} 
    sup[L_{f}{h(x)}+\alpha(h(x))] \geq 0, \forall{x} \in D
\end{equation}
As stated in \cite{MDIC2021}, we also let $C$ = $D$ in this paper.
Next, we recall the following theorem due to \cite{WANG2017}.
\newline
\newline
\textbf{Theorem 2:} \emph{For all time $t$ and all initial state $x_0 \in C$, if the state $x(t,x_0)$ is in set $C$ from Eq. \ref{on_in_bound}-\ref{in_bound}, then set $C$ is forward invariant.}

Next, we define the set $K_{cbf}$ as below,
\begin{equation}
\label{kcbf} 
    K_{cbf}(x) = \{u \in U : L_{f}h(x) + \alpha(h(x)) \geq 0 \}
\end{equation}
Thus, considering the continuous differentiable function h and considering the set $C$ defined by Eq. \ref{on_in_bound}-\ref{in_bound}, if the function $h$ is a control barrier function on the set $C$, according to \textbf{Theorem 2}, for any Lipschitz continuous input $u(x) \in K_{cbf}(x)$ which maps $C$ to $U$, set $C$ is a forward invariant.

In conclusion, using $u$ from Eq. \ref{kcbf} will guarantee that the state will not deviate from set $C$. 

As we will apply the control barrier function on top of coverage control, which will produce $u_{nom}$, we can solve Eq.\ref{unom}-\ref{condition} to attain minimum change of $u$ that satisfy the constraint in the control barrier function.
\begin{eqnarray}
\label{unom}
u = \argmin_u||u-u_{nom}||^2  \\
\label{condition}
s.t.\hspace{0.3cm} L_{f}h(x)+\alpha(h(x)) \geq 0
 \end{eqnarray}
We can rewrite the above constraint as 
\begin{equation}
\label{constraint_new} 
    \dot{h}(x)+\alpha(h(x)) \geq 0
\end{equation}
\section{PROBLEM SETUP}
\begin{figure}[ht]

\centering

\includegraphics[width=0.8\linewidth,clip]{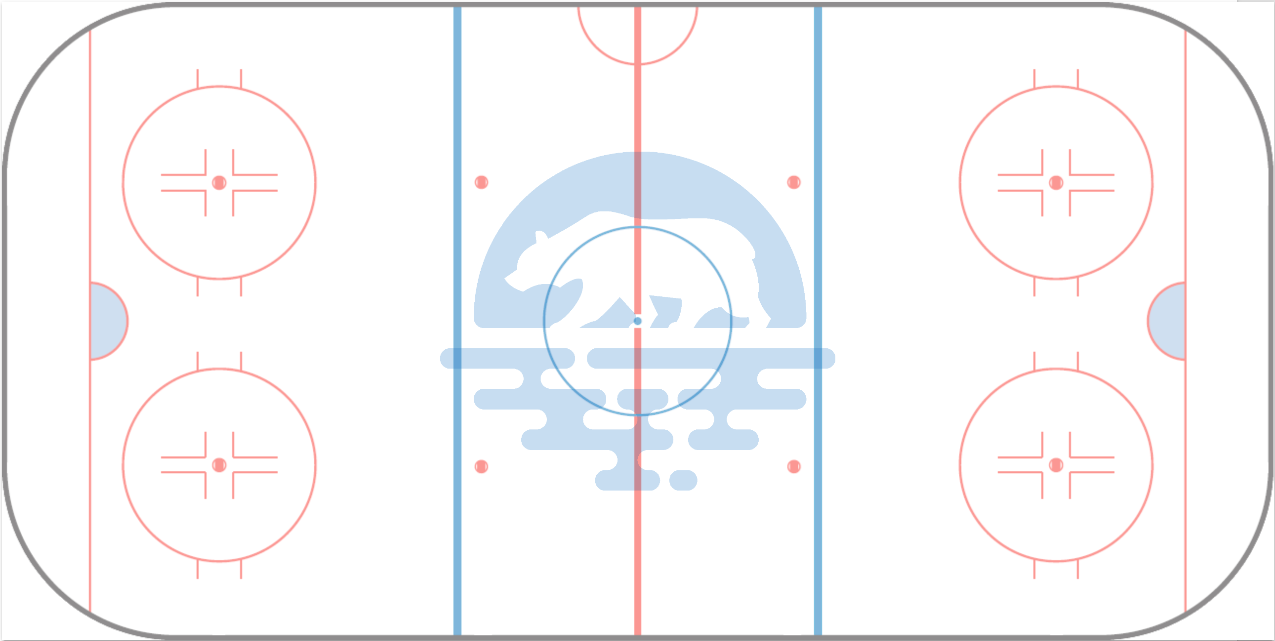}
\caption{Field}
\label{field}
\end{figure}

In this chapter, we describe the general framework of the problem. We consider a defensive phase against an attack by the opposing team in the hockey field, shown in Fig. \ref{field}, which covers 61 m wide and 30 m long. In this paper, we only consider the defensive formation to be only on the left-hand side of the field for simplicity in our simulation.

By inputting the position of the opposing team's players, which is extracted from real scenes, and designing appropriate weight functions and constraints in the control barrier function, we generate the defensive formation of five players, excluding the goalkeeper. We denote $i=1,2,\dots, 5$ as the index of each defensive player,$X_i$ as the position of player $i$, whose dynamics follow Eq. \ref{dynamics}. Moreover, the maximum value of $||u_i||$ is constrained to 3.0 m/s, taking into account the movement speed of an actual ice hockey player. 

Under the above setup, the velocity input $u_i$ of each player $i$ is fixed to the Eq. \ref{control_input}. Once the density function $\phi$ and the constant $k$ is determined, the motion of the defensive players can be simulated. In this paper, we consider the problem of designing a density function $\phi$ and the control barrier function's constraints, that generate defensive formation in general scenes and call this the ice hockey defensive formation generating problem.

\section{CONTROL LOGIC FOR DEFENSIVE MOTION GENERATION}
In the following, we describe the specifications for the control logic qualitatively. The detailed implementation of these specifications will be discussed in the next chapter. 
\subsection{Weight function design}
We first recall the motivation to use coverage control for defense formation. Coverage control is used to make each agent, a defensive player, cover areas on the field, prioritizing significant regions. We can assign the significance of the area by designing the weight function. An area with a higher weight value implies higher importance. In our case, the defensive players should block the offensive players and try to protect the area in front of the goal. 

For blocking the offensive players,
defensive players should stay between offensive players and the goal while also moving according to the offensive players' movement. Accordingly, we would design the weight function to be high where the defensive players should perform blocking. We will also consider the goal as another offensive player to make the defensive player move to protect the goal.

\begin{figure}[ht]

\centering

\includegraphics[width=0.4\linewidth,clip]{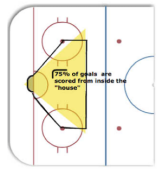}
\caption{House area from \cite{OOTA2020}} 
\label{inhouse_info}
\end{figure}

For covering the significant regions, the area closer to the goal should be more significant than the one further away. Moreover, the area in front of the goal should also be significantly important. Specifically, defensive players should protect the yellow area, which is shown in Fig.\ref{inhouse_info}. According to \cite{OOTA2020}, this area contributes to 75\% of goals scored in a real game.

Next, the defensive players should pay more attention to the puck holder, which is the player who holds the puck. When the puck holder moves closer to the goal, we can also partially ignore the vertically opposite side of the field, omitting the goal's weight on the vertical-opposite side but still paying attention to offensive players in the area.

\begin{figure}[ht]

\centering

\includegraphics[width=0.4\linewidth,clip]{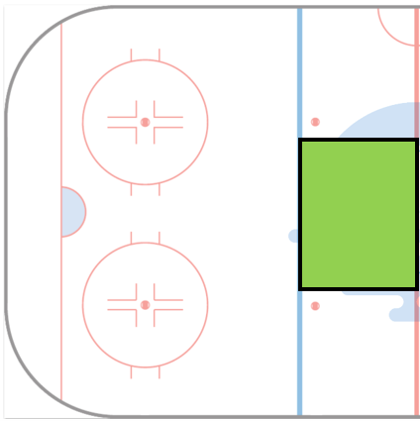}
\caption{Area to lower the gain} 
\label{blue_field}
\end{figure}

Lastly, the middle area on the right side of the blue line empirically should also have lower importance. We can visualize that area in Fig.\ref{blue_field}.

\subsection{CBF design}
The control barrier function is used in this problem to simulate goal block and pass cut movements according to \cite{MDIC2021}. In our simulation, we choose to focus only on performing a pass cut, which is the action of a defensive player to move in between two offensive players. In the previous CBF model in \cite{MDIC2021}, a line-based CBF is proposed for realizing pass cut. However, this model does not guarantee to simulate such action. 
\subsubsection{Drawbacks of line-based CBF}
In the previous model from \cite{MDIC2021}, an orthogonal distance toward a line, running through two offensive players' positions, is used as a model for a control barrier function for simulating a pass cut. In this case, CBF is non-negative when the defensive player is no further than $\delta$ from the line, as shown in Eq.\ref{old_pass_cut_cbf}.
\begin{equation}
\label{old_pass_cut_cbf} 
    h_{i}^{cut} = \delta^2 - \dfrac{(aX_{x}-X_{y}+b)^2}{a^2+1}
\end{equation}
where $[X_{x} , X_{y}]^T$ is the position of the defensive player. Let other variables in Eq. \ref{old_pass_cut_cbf} are defined as 
\begin{equation}
\label{a}
a = \dfrac{X_{y1}-X_{y2}}{X_{x1}-X_{x2}}  
\end{equation}
\begin{equation}
\label{b}
b = \dfrac{X_{x1}X_{y2}-X_{x2}X_{y1}}{X_{x1}-X_{x2}}
 \end{equation}
where $[X_{xk} , X_{yk}]^T$ is the position of the $k^{th}$ offensive player for $k \in {1,2}$.

\begin{figure}[ht]
\centering
\includegraphics[width=0.6\linewidth,clip]{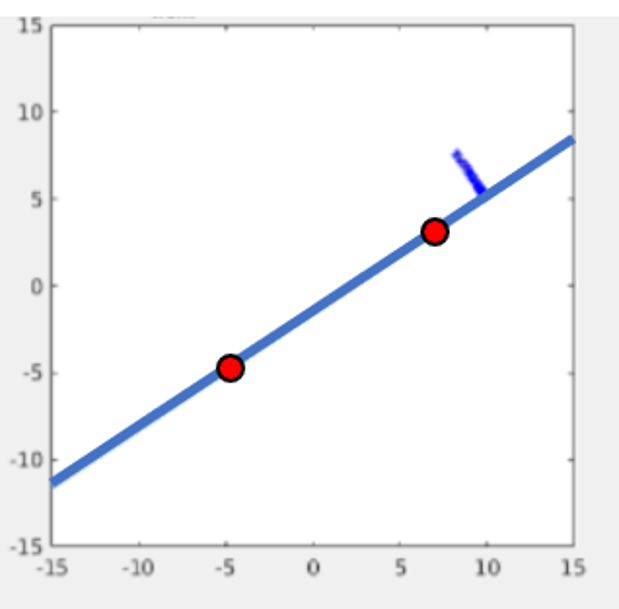}
\caption{Improper result using line-based CBF}
\label{line_cbf_ncor}
\end{figure}

There are two problems in the previous model. First, Eq. \ref{a} and \ref{b} are incalculable when $X_{x1} = X_{x2}$. Second, this control barrier function doesn't guarantee that the defensive player moves between two offensive players as shown in Fig.\ref{line_cbf_ncor}.  
\begin{figure}[ht]
\centering
\includegraphics[width=0.6\linewidth,clip]{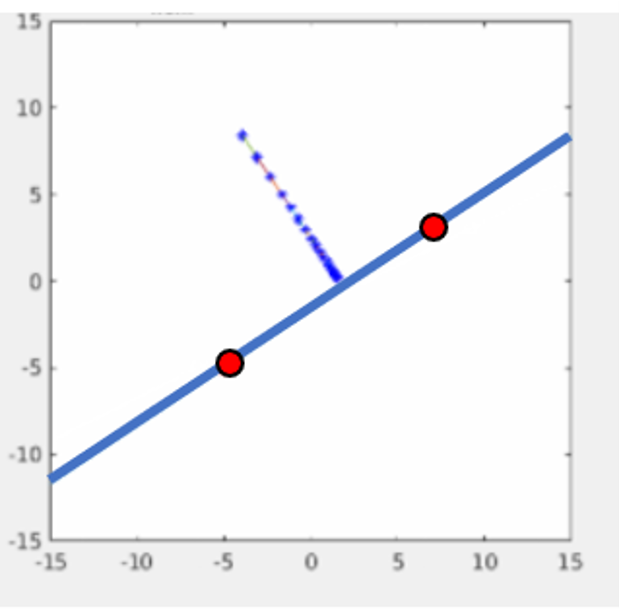}
\caption{Proper result using line-based CBF}
\label{line_cbf_cor}
\end{figure}

Fig. \ref{line_cbf_ncor} demonstrates the incorrect movement as the blue star dot, representing defensive player, does not move in between two red dots, representing static offensive players. This happens because the line, defined from two offensive players, stretches beyond the positions of two.

\subsubsection{Ellipsoidal CBF}
\begin{figure}[ht]
\centering
\includegraphics[width=0.6\linewidth,clip]{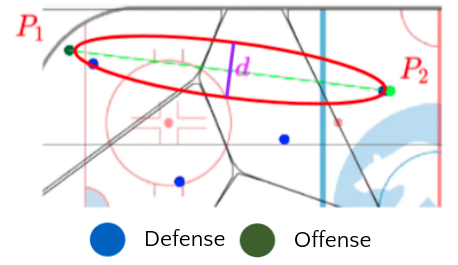}
\caption{Example of ellipsoidal control barrier function}
\label{ellipsoid}
\end{figure}

 We propose to overcome this problem by introducing the ellipsoidal model as a control barrier function. The motivation of using the ellipsoidal model is to force the defensive players to move inside the ellipsoid, defined by the position of two offensive players, as shown in Fig.\ref{ellipsoid}.

\section{IMPLEMENTATION OF CONTROL LOGIC}
In this chapter, we describe the implementation of the control logic according to the specifications in the previous chapter. At the end of the chapter, we also describe the comprehensive algorithm used in the simulation.

\subsection{Weight function implementation}
We assign a weight, described in chapter 2, to each offensive player and the goal by applying a two-dimensional multivariate Gaussian distribution as a weight to each position. When we decide the position to apply the distribution for each player, we also consider their velocity and the nature of the defensive player, which tends to stay between offensive players and the goal, as stated in the previous chapter. We notate the position of the weight that represents offensive player $i$ as $P_{wi}$. The weight position can be described as shown in the equation below.
\begin{equation}
\label{gaussian} 
    P_{wi} = P_i + \dfrac{P_{goal}-P_i}{|P_{goal}-P_i|} + \dot{P}_i  \
\end{equation}
where $P_i$ is $i^{th}$ offensive players' positions and $P_{goal}$ is the position of the goal which is $[6,15]^T$. After determining the position to apply the Gaussian distribution, we use the following covariance matrix as a specification for applying the distribution to each position defined above.
 \begin{equation}
    \label{convarience}
     \Sigma_{Wi} = \left(
    \begin{array}{ccc}
      15 & 0  \\
      0 & 15 
    \end{array}
  \right)  \\
 \end{equation}
 In conclusion, we have n+1 weight distributions on the field which are $W_1,W_2,..,W_n$ and $W_{goal}$.

\begin{figure}[ht]

\centering

\includegraphics[width=0.8\linewidth,clip]{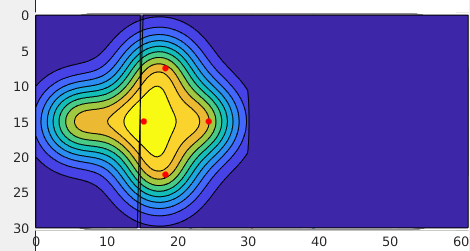}
\caption{Weight assigned for each offensive player and at the goal position}
\label{weight_pos}
\end{figure}

 Once applied the distribution as a weight to each position stated above, the result can be seen in Fig.\ref{weight_pos}, given that the red dots are offensive players.

\begin{figure}[ht]

\centering

\includegraphics[width=0.8\linewidth,clip]{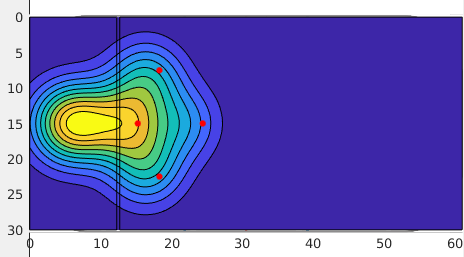}
\caption{Weight from Fig.9 when applying distance gain} 
\label{distance field}
\end{figure}

Next, we apply distance gain to each of the weights representing an offensive player, prioritizing the area closer to the goal with a higher weight value. We design the distance gain to have a higher value when applying to the position closer to the goal to achieve this specification. The updated weights are
\begin{equation}
\label{distance_gain} 
\scalemath{0.8}{
    W^*_i(x,y) = W_i(x,y) * \dfrac{(25 - |P_{goal} - [x,y]^T|)}{25}, \forall{i} \in {1,2,...,n},\forall{[x,y]^T} \in Q
}
\end{equation}
where $W^*_i(x,y)$ is the updated weight from the $i^{th}$ weight distribution at the position (x,y) on the field $Q$. Fig.\ref{distance field} demonstrates the result of the updated weights.

\begin{figure}[ht]

\centering

\includegraphics[width=0.8\linewidth,clip]{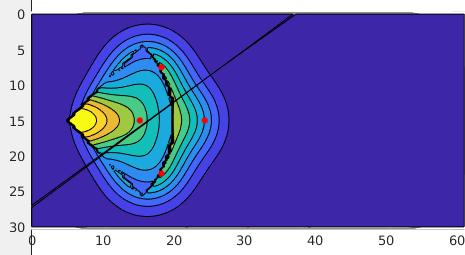}
\caption{Weight from Fig.10 when applying gain on house area ($p=2$)} 
\label{inhouse_weight}
\end{figure}

Furthermore, another area in the field which defensive players should protect is the house area, as stated in the previous chapter. Therefore, we also imitate the house area and apply constant gain to raise the weights in the area. We design the house area as the inner area bounded by Eq. \ref{l1}-\ref{circle_house}.

\begin{eqnarray}
\label{l1}
y_1 &= -x_1 + 20  \\
\label{l2}
y_2 &= x_2 + 10 
\end{eqnarray}
\begin{equation}
\label{circle_house}
(x_2-5)^2+(y_2-15)^2=225
\end{equation}
In this paper, we denote this bounded area as $Y$. We apply a constant gain $p$ to the weight inside this area. Furthermore, we also remove the goal weight outside the house zone. Using the above specification, we define the resulting weight as 

\begin{eqnarray}
\label{distance_yellow} 
    W^{**}_i(x,y) &=& W^*_i(x,y) * p, \forall{(x,y)} \in Y \\
    &&\forall{i} \in {1,...,n} \nonumber \\
    W^{**}_{goal}(x,y) &=& \begin{cases}
    W^*_{goal}(x,y) * p * 1.5  &\text{$\forall{(x,y)}$ $\in$ Y \nonumber}\\
    0 &\text{otherwise}
    \end{cases} 
\end{eqnarray}
where $Y$ is the inner area described by Eq. \ref{l1}-\ref{circle_house} and $p$ is the gain constant. By setting $p=2$, the resulting weight can be seen in Fig. \ref{inhouse_weight}.

To prioritize the puck holder when the puck holder moves close to the goal area, we selectively apply a constant gain to lower the significance of other offensive players on the field as shown in Alg. \ref{alg:the_alg1}. We denote the puck holder's index as $pu$ $\in \{1,2,...,n\}$ and the x and y position of the puck holder as $P_{pu\_x}$ and $P_{pu\_y}$ respectively.

\begin{algorithm}
\caption{Weight adjustment algorithm}
\label{alg:the_alg1}
\hspace*{\algorithmicindent} \textbf{Result} $W^{***}_i(x,y) \hspace*{0.5cm} \forall{i} \in \{1,2,...,n,goal\}$, a set of weights. 
\begin{algorithmic}[1]
\small
\If{$P_{pu\_x} \leq 10$ and $P_{pu\_y} \leq 15$} 
    \For{$i \gets 1$ to $n$}      
        \If{i != J}
            \State $W^{***}_i(x,y) = 0.1 * W^{**}_i(x,y)$ $\forall{y}  <$ 15
            \State $W^{***}_i(x,y) = 0.05 * W^{**}_i(x,y)$ $\forall{y}  \geq$ 15
        \EndIf
    \EndFor
    \State $W^{***}_{goal}(x,y) = W^{**}_{goal}(x,y)$ \hspace*{0.1cm} $\forall{y}  <$ 15 
    \State $W^{***}_{goal}(x,y) = 0$ \hspace*{0.1cm} $\forall{y} \geq$ 15 or $x \geq$ 10 
\ElsIf{$P_{pu\_x} \leq 10$}
    \For{$i \gets 1$ to $n$}      
        \If{i != J}
            \State $W^{***}_i(x,y) = 0.1 * W^{**}_i(x,y)$ $\forall{y} \geq$ 15
            \State $W^{***}_i(x,y) = 0.05 * W^{**}_i(x,y)$ $\forall{y} <$ 15
        \EndIf
    \EndFor
    \State $W^{***}_{goal}(x,y) = W^{**}_{goal}(x,y)$ \hspace*{0.1cm}$\forall{y} >$ 15 
    \State $W^{***}_{goal}(x,y) = 0$ \hspace*{0.1cm} $\forall{y} \leq$ 15 or $\forall{x} \geq$ 10 
\EndIf 

\end{algorithmic}
\end{algorithm}

\begin{figure}[ht]

\centering

\includegraphics[width=0.8\linewidth,clip]{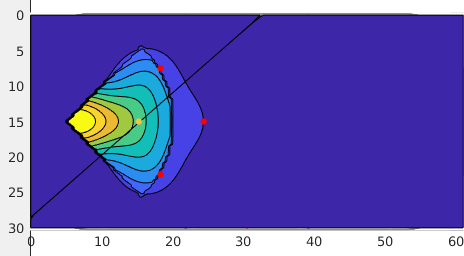}
\caption{Weight from Fig.11 when applying constant gain.} 
\label{p_gain_weight}
\end{figure}

Given that the orange dot represents the puck holder, the result of the above weights can be visualized as Fig.\ref{p_gain_weight}.

Empirically, the green area in Fig. \ref{field} is also less important than other areas. As a result, we apply a constant gain to lower this specific area. We can define the updated weight by applying the following equation 

\begin{eqnarray}
\label{blue_gain} 
\scalemath{0.8}{
    W^{****}_i(x,y) = \begin{cases}
W^{***}_i(x,y) * 0.1  &\text{$\forall{x} \in$ [14.945,21.960]}\\
&\text{and $\forall{y} \in$ [10,20]}\\
W^{***}_i(x,y) &\text{otherwise}

\end{cases} 
} 
\end{eqnarray}
\begin{equation*}
\scalemath{0.7}{
    W^{****}_{pu}(x,y) = W^{***}_{pu}(x,y)
    }
\end{equation*}
\begin{equation*}
\scalemath{0.7}{
    W^{****}_{goal}(x,y) = W^{***}_{goal}(x,y)
    }
\end{equation*}
for i $\in$ $\{1,2,...n\}\symbol{92} pu$. In other words, this specification only applies to non-puck holder offensive players.

Lastly, as we focus only on the left-hand side of the field, we omit all the weight on the right side of the field.

\begin{equation}
\label{final_weight}
\scalemath{0.8}{
    W^{final}_i(x,y) = \begin{cases}
0  &\text{$\forall{x}$ $>$ 30}\\
W^{****}_i(x,y) &\text{otherwise} \end{cases} 
\forall{i} \in {1,2,..,n,goal}
}
\end{equation}
This is the finalized weights that we will use in the simulation.
\subsection{Ellipsoidal CBF implementation}
As stated in section 5.1, we design the new CBF as the equation below.
\begin{equation}
\label{ellipsoidal model}
    h(X,X_1,X_2) = 1-(X-X_o)^{T}P(X-X_o)
 \end{equation}
 where
\begin{equation}
\label{P model}
    P = 
        \scalemath{0.7}{
            \left[
                \begin{array}{cc} \left( \dfrac{\rm{cos}^2\theta}{(l/2)^2} + \dfrac{\rm{sin}^2\theta}{(d/2)^2} \right) &  \dfrac{\rm{sin}2\theta}{2} \left( \dfrac{1}{(l/2)^2} - \dfrac{1}{(d/2)^2} \right) \\ \dfrac{\rm{sin}2\theta}{2} \left( \dfrac{1}{(l/2)^2} - \dfrac{1}{(d/2)^2} \right)  & \left( \dfrac{\rm{sin}^2\theta}{(l/2)^2} + \dfrac{\rm{cos}^2\theta}{(d/2)^2} \right) \end{array} 
        \right]
        }
 \end{equation}
 given that $X = [x,y]^T \in Q$ is the position of the defensive player, $X_o  = [x_o,y_o]^T \in Q$ is the center of the ellipse formed by having the position of offensive two players $X_1, X_2$ as vertices and the height of the semi-minor axes $d$, and $X_k$ = $[x_k,y_k]^T$ is the position of the $k^{th}$ offensive players for $k \in {1,2}$.
 
 Furthermore, the parameters in Eq. \ref{P model} can be defined as the following.
 \begin{equation}
 \label{x0 equation}
        X_o = \dfrac{X_1 + X_2}{2}, l = ||X_2 - X_1||
\end{equation}
where $l$ is the width of the semi-major axes and $ \theta $ is defined as below
\begin{eqnarray}
\label{theta equation}
\theta &=& \rm{arctan}\left( \dfrac{y_2 - y_1}{x_2-x_1} \right),  \\
\label{cos equation}
\rm{cos}(\theta) &=& \dfrac{x_2-x_1}{l}, \\
\label{sin equation}
\rm{sin}(\theta) &=& \dfrac{y_2-y_1}{l}
\end{eqnarray}

With the above ellipsoidal model, $h$ will be non-negative only when the defensive player moves inside the ellipse, between 2 offensive players.

As, we define the new ellipsoidal model, we rewrite the resulting constraint as Eq. \ref{H big view} by applying it to Eq. \ref{constraint_new}. 
\begin{equation}
\begin{split}
\label{H big view}
    \dot{h}(X,X1,X2) &= \dfrac{\partial h}{\partial X}\dot{X}+\dfrac{\partial h}{\partial X_1}\dot{X_1}+\dfrac{\partial h}{\partial X_2}\dot{X_2} \\ &= \dfrac{\partial h}{\partial X}u+\dfrac{\partial h}{\partial X_1}\dot{X_1}+\dfrac{\partial h}{\partial X_2}\dot{X_2}\\ & \geq -\alpha(h(X,X_1,X_2))
\end{split}
\end{equation}
The derivative of h can be expanded into the following equation.
\begin{equation}
\begin{split}
\label{H expand}
    \dot{h}(X,X1,X2) &= a(\dot{x_1}+\dot{x_2}-2\dot{x})\\
                     &+ b(\dot{x_1}-\dot{x_2})\\
                     &+ c(\dot{y_1}+\dot{y_2}-2\dot{y})\\
                     &+ d(\dot{y_2}-\dot{y_1})
\end{split}
\end{equation}
where a,b,c and d are 
\begin{equation}
\begin{split}
\label{a_h}
    a &= \Bigg[ \left(\dfrac{\rm{cos}^2(\theta)}{(l/2)^2} + \dfrac{\rm{sin}^2(\theta)}{(d/2)^2} \right)(x-x_0) \\ &+ \rm{cos}\theta \rm{sin} \theta \left( \dfrac{1}{(l/2)^2} - \dfrac{1}{(d/2)^2} \right)(y-y_0) \Bigg]
\end{split}
\end{equation}
\begin{equation}
\begin{split}
\label{b_h}
    b &= \Bigg[ - \bigg( \left( \dfrac{1}{(d/2)^2} - \dfrac{1}{(l/2)^2}\right) \rm{sin}(2\theta) \left( \dfrac{y_2 - y_1 }{l^2} \right) \\ &+ \dfrac{\rm{cos}^2(\theta) (x_1 - x_2)}{2(l/2)^4} \bigg)(x-x_0)^2 \\ &- 2\bigg(\rm{sin} (2\theta) \left( \dfrac{4(x_2-x_1)}{l^4} \right) \\
&+ \bigg( \dfrac{1}{(l/2)^2} - \dfrac{1}{(d/2)^2} \bigg) \rm{cos}(2\theta) \left( \dfrac{y_2 - y_1}{l^2} \right) \bigg)\\ &(x-x_0)(y-y_0) \\
&- \bigg( \left( \dfrac{1}{(l/2)^2} - \dfrac{1}{(d/2)^2}\right) \rm{sin}(2\theta) \left( \dfrac{y_2 - y_1 }{l^2} \right) \\ &- \dfrac{\rm{sin}^2(\theta) (x_1 - x_2)}{2(l/2)^4}   \bigg)(y-y_0)^2 \Bigg] 
\end{split}
\end{equation}
\begin{equation}
\begin{split}
\label{c_h}
    c &= \Bigg[\rm{cos}\theta \rm{sin} \theta \left( \dfrac{1}{(l/2)^2} - \dfrac{1}{(d/2)^2} \right)(x-x_0) \\
&+ \left( \dfrac{\rm{sin}^2(\theta)}{(l/2)^2} + \dfrac{\rm{cos}^2(\theta)}{(d/2)^2} \right) (y-y_0) \Bigg]
\end{split}
\end{equation}
\begin{equation}
\begin{split}
\label{d_h}
    d &= \Bigg[ -\bigg(\left( \dfrac{1}{(d/2)^2} - \dfrac{1}{(l/2)^2}\right) \rm{sin}(2\theta) \left( \dfrac{x_2 - x_1 }{l^2} \right) \\ & - \dfrac{\rm{cos}^2(\theta) (y_2 - y_1)}{2(l/2)^4} \bigg)(x-x_0)^2 \\ &- 2\bigg( \rm{sin} (2 \theta) \left( \dfrac{4(y_1-y_2)}{l^4} \right) \\ & + \left( \dfrac{1}{(l/2)^2} -  \dfrac{1}{(d/2)^2} \right) \rm{cos} (2\theta) \left( \dfrac{x_2 - x_1}{l^2} \right) \bigg) \\ & (x-x_0)(y-y_0) \\ &-
 \bigg( \left( \dfrac{1}{(l/2)^2} - \dfrac{1}{(d/2)^2}\right) \rm{sin}(2\theta) \left( \dfrac{x_2 - x_1 }{l^2} \right) \\ &- \dfrac{\rm{sin}^2(\theta) (y_2 - y_1)}{2(l/2)^4} \bigg)(y-y_0)^2 \Bigg]
\end{split}
\end{equation}
Lastly, in our simulation, $d$ is set to 0.01, and the $\alpha$ function in the constraint equation is set to the h function. For performing pass cut, $P_2$ will always be puck holder. Thus, we can consider that the ellipsoid CBF only needs 2 inputs which are the position of defensive player $X$ and the position of another offensive player which should be blocked $X_1$.
\subsection{Online selection of CBFs}
In the following, we explain the algorithm used in the simulation to selectively choose the pairs of a defensive player and a offensive player to apply CBF to perform a pass cut. 

Pass cut is an action of blocking the player from passing the puck to other players. Alg.\ref{alg:selective_alg} is designed based on this motivation. Specifically, we aim to apply CBF when there is a possibility that offensive players, who are inside the house area, can receive the puck and subsequently score a goal. Therefore, we design an algorithm to only select defensive players inside the house area to perform a pass cut and prioritise block offensive players closer to the goal than those further away.

 \begin{algorithm}[ht]

\caption{Online selection CBF algorithm}
\label{alg:selective_alg}
\hspace*{\algorithmicindent} \textbf{Result} $P_{passcut}$, a set of pairs of defensive player' indexes and offensive player' indexes to perform pass cut. 
\begin{algorithmic}[1]

\State{Identify offensive players except puck holder in the house area as $O_{house}$ and defensive players in the house area as $D_{house}$.}
\State{Let $P_{passcut}$ = $\varnothing$}
\While{$O_{house} \neq \varnothing $ and $D_{house} \neq \varnothing $ }

\State{Select $i \in O_{house} = \argmin\limits_{i} ||x_i-x_{goal}||$ O, identify $d_{i} \in D$ = $\argmax\limits_d$ h($P_d,P_i,P_{puck}$)}
\State{Add $(d_{i},i)$ to $P_{passcut}$}
\State{Remove $d_i$ from $D_{house}$ and $i$ from $O_{house}$}
\EndWhile
\end{algorithmic}
\end{algorithm}

From Alg.\ref{alg:selective_alg}, we compute the pairs of offensive players that need to be blocked and the corresponding defensive players to perform pass cut. In the next section, we will apply the ellipsoidal CBF to each pair in the set $P_{passcut}$, preventing offensive players in the house area from receiving the puck from the puck holder. 

\subsection{Comprehensive algorithm}
\begin{algorithm}[ht]
\caption{Comprehensive algorithm}
\label{alg:comprehensive_alg}
\hspace*{\algorithmicindent} \textbf{Result} $u$, a set of inputs for defensive players $\{1,2,...,5 \}$
\begin{algorithmic}[1]
\State{Initialize position of defensive players i $\in$ $D = \{1,2,..,5\}$ as $X_{di} \in Q$}
\State{Let the position of offensive player j $\in$ $O = \{1,2,...,n\}$ as $X_{oj} \in Q$}
\State{Let the weight function as $\phi(x,y)$ with the design and implementation stated in section 5.1 using $X_{oj}$}
\State{Compute Voronoi cell $C_i$ using $X_{di}$ and $\phi(x,y)$ in Eq. \ref{voronoi}}
\State{Compute cent($C_i$) using $C_i$ and $\phi(x,y)$ in Eq. \ref{cent_voronoi} $\forall{i}$ $\in D$ }
\State{Compute $u^{nom}_i$ using $\rm{cent}(C_i)$ and $\phi(x,y)$ in Eq. \ref{control_input} with $k$ = 1 $\forall{i}$ $\in D$ }
\State{Let $P_{passcut}$ is the output of Algorithm 2.}
\For{$i \gets 1$ to $5$}      
        \If{i is the first index of pair $p \in$ $P_{passcut}$}
            \State{Let $k$ = second index of the pair $p$}
            \State Compute $u_i$ by applying the ellipsoidal CBF with $x_i$ and the corresponding $x_k$ using quadratic programming to solve Eq. \ref{condition} subject to Eq. \ref{constraint_new}..
        \Else
            \State $u_i$ = $u^{nom}_i$
        \EndIf
        \If{$||u_i||$ $>$ 3}
            \State $u_i$ = $\dfrac{u_i}{||u_i||}*3$
        \EndIf
    \EndFor

\end{algorithmic}
\end{algorithm}

In the following, we describe the comprehensive algorithm in Alg.\ref{alg:comprehensive_alg}, which is used to calculate $u_i$ for each offensive player by applying the theorem, specifications, and algorithms stated above.

After computing the input for each defensive player $u_i$ from Alg.\ref{alg:comprehensive_alg}, we input the control inputs to the system and update the position of defensive players. Then, we do this process again with the new offensive player's positions in the following frame until the simulation ends.

\section{SIMULATION STUDY}
In this chapter, we explain the method to attain the data from real scenes in this papar. Then, we describe the evaluation method that we will use to evaluate the simulation results. Lastly, we show the results of the simulation using the real scenes' data.
\subsection{Data extraction from real scene}
\begin{figure}[ht]

\centering

\includegraphics[width=1\linewidth,clip]{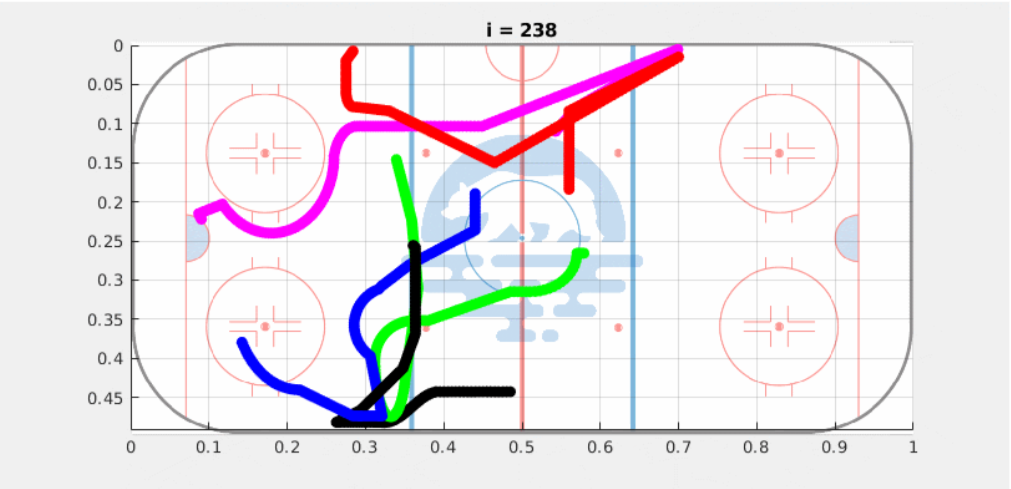}
\caption{Example of extracted data from real scenes}
\label{extract_data}
\end{figure}

For the data, we use ice hockey game scenes from Winter Olympic 2018. We selected two defensive scenes in which we can track offensive players almost all the time to attain the most accurate input for the simulation. The scenes are chosen among several scenes proposed by Takahito Suzuki, a retired head coach of the Japanese national team with who we cooperate within this research. The scenes used in this simulation are as the following:
\begin{itemize}
  \item Canada and Czech republic's game 
  \item Canada and Finland's game
\end{itemize}
For extracting the data from the scene, we have tried techniques according to \cite{NEIL2020}. However, we discovered two problems using these techniques. First, we are not able accurately track players' poses using the pose estimation tool due to frequent collisions between players. Second, the court parsing process can not track the template images, needed to extract camera movement from these particular scenes because the number of the template images in these specific scenes is not high enough. As a result, we decide to extract the data manually. We extract the data frame by frame using the human eye to observe the position and movement of the players on the scenes. The example of the extracted data from Canada and Finland's game scene can be seen in Fig. \ref{extract_data}.

From figure \ref{extract_data}, each coloured line represents one offensive player's motion throughout the scene.
\subsection{Evaluation Method}
After surveying many papers and journals related to the evaluation of ice hockey defensive strategy, we found that there are no evaluation metric which can assess how well the generated formation is. The current evaluation metrics are based on statistical data such as \cite{STAT2017} and \cite{STAT2018}. Thus, we decide to use empirical method to evaluate our result by collaborating with Takahito Suzuki to evaluate the generated formations. 

\subsection{Results}
This section shows results from simulating the defensive formation using the real data previously mentioned in this chapter. We use blue dots to represent defensive players, an orange dot to represent a puck holder, and red dots to represent other offensive players. This section also shows the simulation results using different $p$ values defined in Eq.\ref{distance_yellow}. 

\begin{figure}[ht]

\centering

\includegraphics[width=0.6\linewidth,clip]{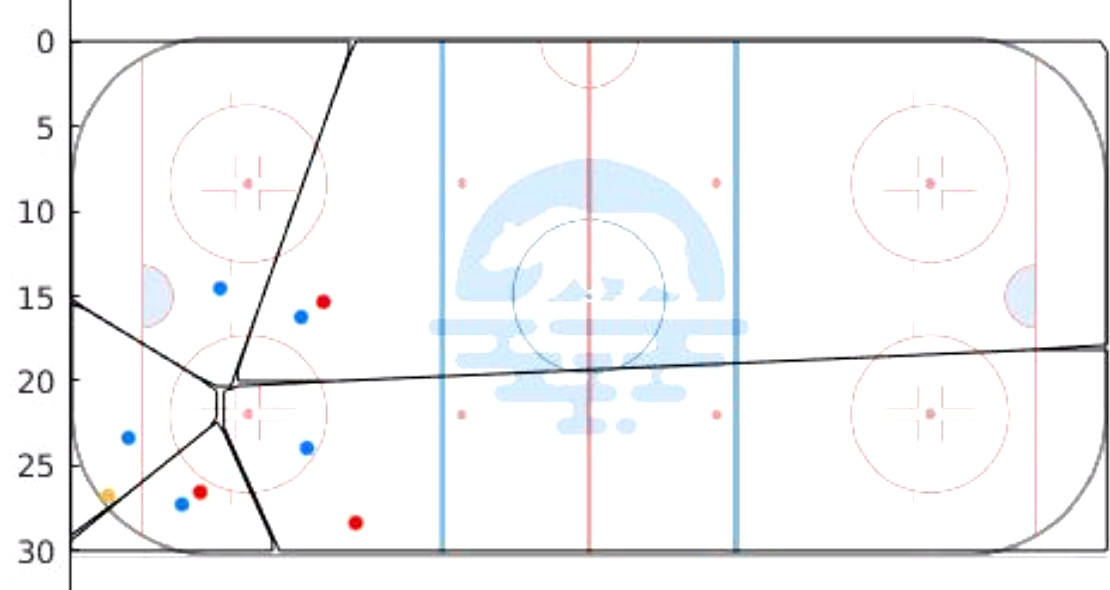}
\caption{The screenshots of the simulation from Canada and Czech republic's game 1} 
\label{Canada and Czesh1}
\end{figure}

\begin{figure}[ht]

\centering

\includegraphics[width=0.6\linewidth,clip]{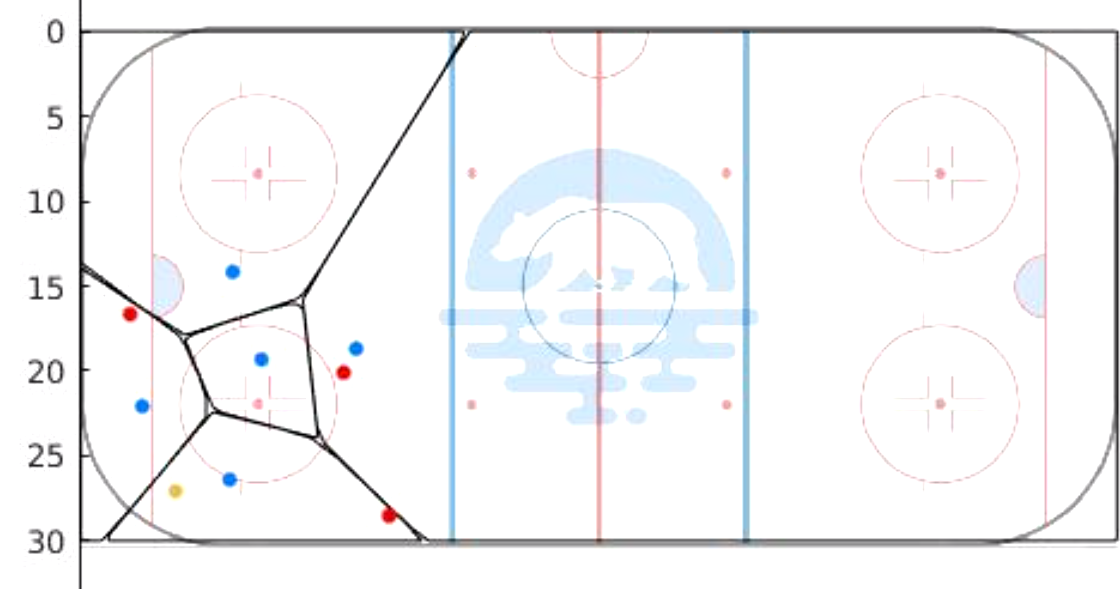}
\caption{The screenshots of the simulation from Canada and Czech republic's game 2} 
\label{Canada and Czesh2}
\end{figure}

\begin{figure}[ht]

\centering

\includegraphics[width=0.6\linewidth,clip]{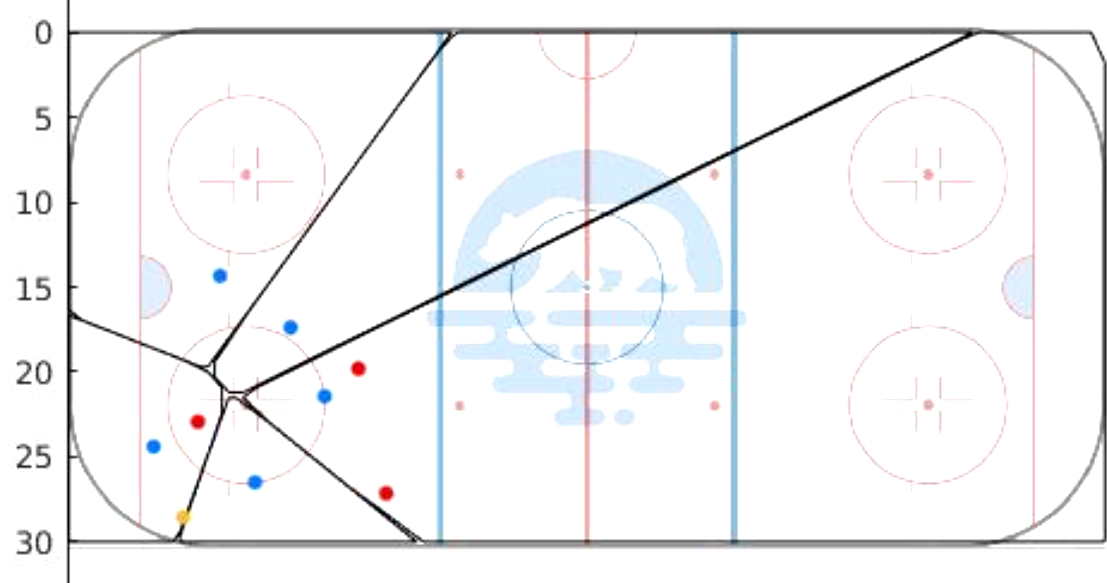}
\caption{The screenshots of the simulation from Canada and Czech republic's game 3} 
\label{Canada and Czesh3}
\end{figure}

The screenshots of the simulation from Canada and Czech republic's game using $p$ = 2 can be seen in Fig. \ref{Canada and Czesh1}-\ref{Canada and Czesh3}.

\newpage

\begin{figure}[ht]

\centering

\includegraphics[width=0.6\linewidth,clip]{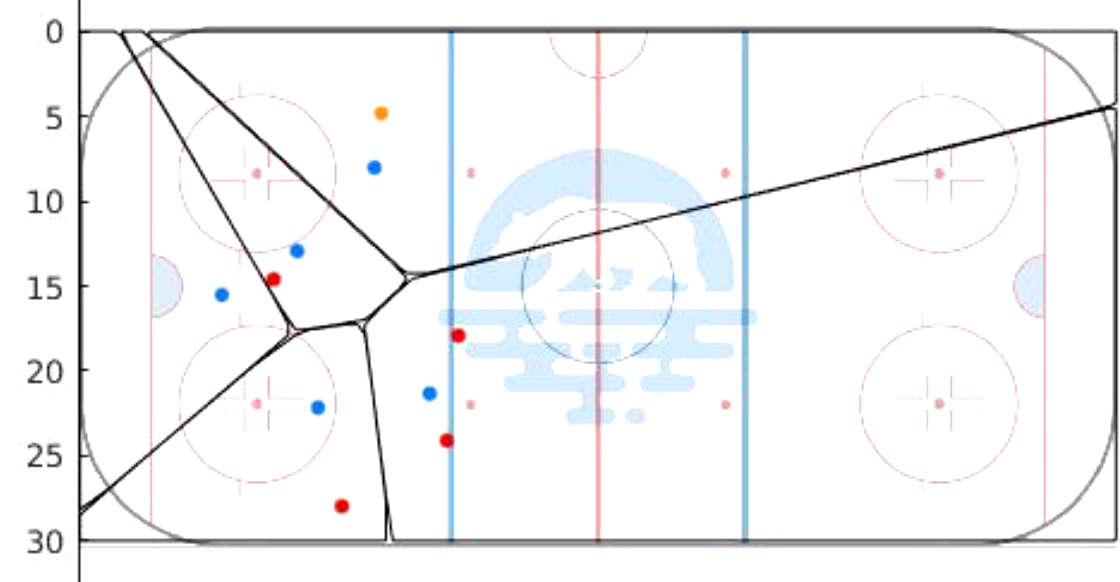}
\caption{The screenshots of the simulation from Canada and Finland republic's game 1} 
\label{Canada and Finland1}
\end{figure}

\begin{figure}[ht]

\centering

\includegraphics[width=0.6\linewidth,clip]{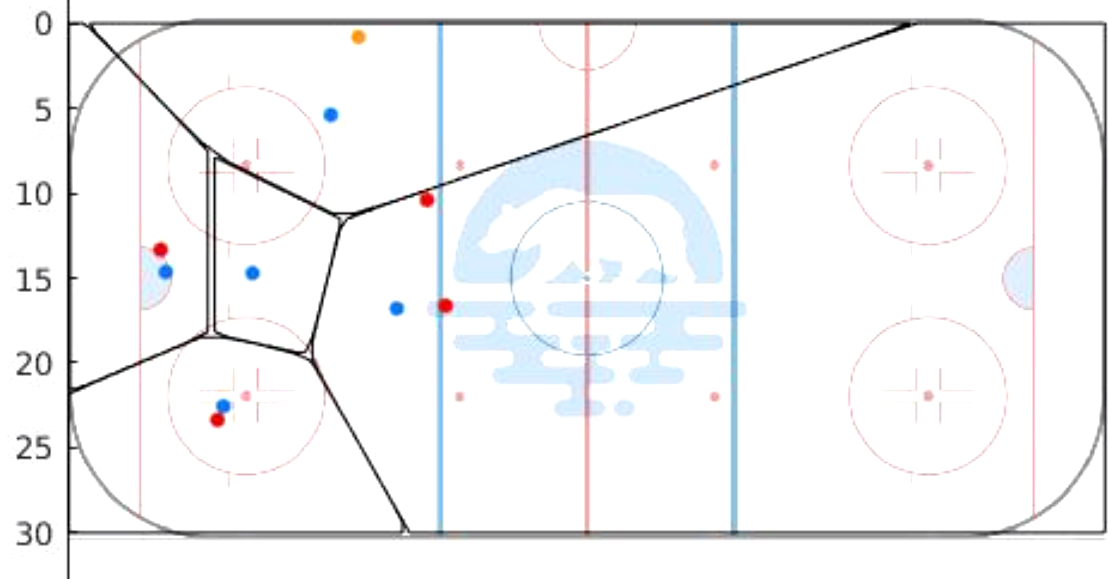}
\caption{The screenshots of the simulation from Canada and Finland republic's game 2} 
\label{Canada and Finland2}
\end{figure}

\begin{figure}[ht]
\centering
\includegraphics[width=0.6\linewidth,clip]{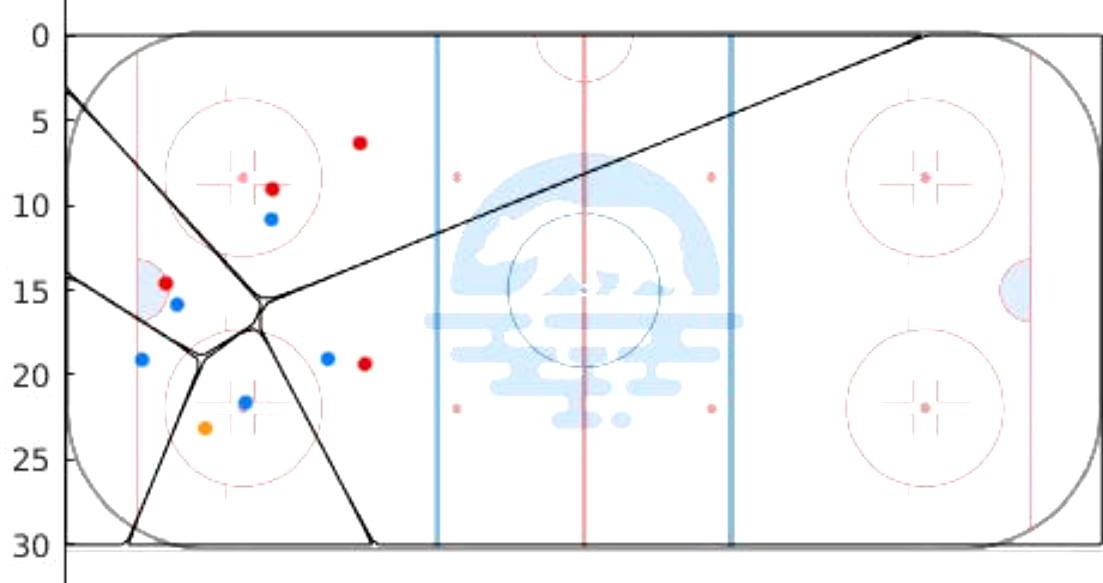}
\caption{The screenshots of the simulation from Canada and Finland republic's game 3} 
\label{Canada and Finland3}
\end{figure}

The screenshots of the simulation from Canada and Finland's game using $p$ = 2 can be seen in Fig. \ref{Canada and Finland1} - \ref{Canada and Finland3}.

Both simulations demonstrate the defensive movement of defensive players defending the area in front of the goal while also moving along with offensive players. These results are also being evaluated by the professional, as mentioned in section 6.2. He assesses that the generated movements of the defensive players are closed to the actual players. Thus, we can conclude that we achieve the generation of defensive formation for the scenes. 

\begin{figure}[ht]

\centering

\includegraphics[width=0.6\linewidth,clip]{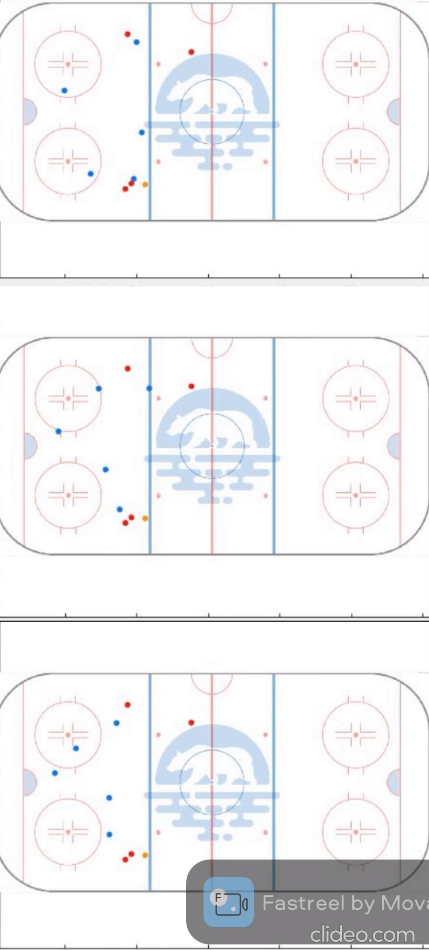}
\caption{The screenshots of the simulation from Canada and Finland's game using different $p$ values (0.5,3 and 10 from top to bottom) 1} 
\label{p val 1}
\end{figure}

\begin{figure}[ht]

\centering

\includegraphics[width=0.6\linewidth,clip]{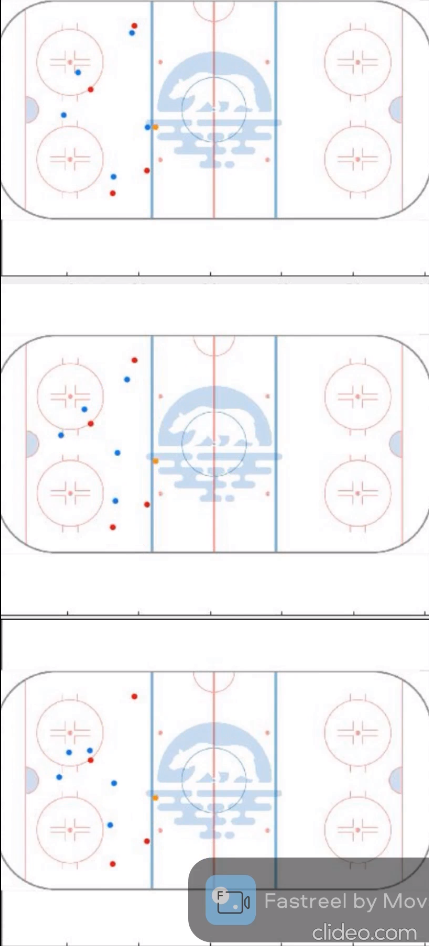}
\caption{The screenshots of the simulation from Canada and Finland's game using different $p$ values (0.5,3 and 10 from top to bottom) 2} 
\label{p val 2}
\end{figure}

\begin{figure}[ht]

\centering

\includegraphics[width=0.6\linewidth,clip]{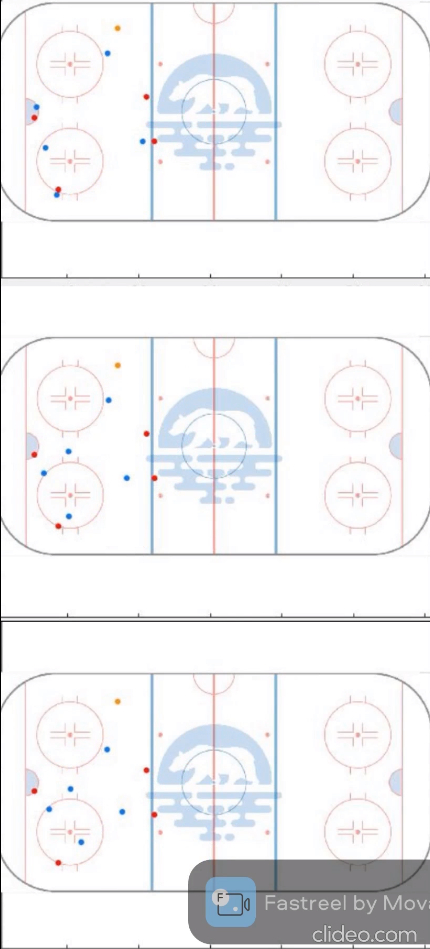}
\caption{The screenshots of the simulation from Canada and Finland's game using different $p$ values (0.5,3 and 10 from top to bottom) 3} 
\label{p val 3}
\end{figure}

Furthermore, the results from the simulation using Canada and Finland's game with different $p$ values, which are 0.5,3 and 10 from top to bottom respectively, as shown in Fig.\ref{p val 1}-\ref{p val 3}.

Fig. \ref{p val 1}-\ref{p val 3} show that different $p$ value gives different results. With a lower $p$ value, the importance of the house area is relatively low, and defensive players will move close to each offensive player instead. A low $p$ value leads to movements that are close to a peer-to-peer defensive strategy in the real game. On the other hand, when $p$ grows bigger, the weight on the house area becomes higher, increasing the area's significance. Consequently, defensive players exhibit defensive behavior that resembles zone defense strategy in the real game, where defensive players move to cover the important area rather than the players. We can also observe the intermediate strategies when we set $p = 3$.

\section{CONCLUSION}
In conclusion, we further develop the control logic from \cite{OOTA2020} and \cite{MDIC2021} by changing the weight function's design in coverage control and improving the CBF model to an ellipsoidal CBF. Furthermore, we confirm that we can generate the defensive formation of ice hockey players using a new logic for two general scenes without adjusting the weight function specifically for the scene. We also discover that we can produce different strategies by changing the weight ratio parameter $p$, which affects the ratio between the players' weights and the area's weight in the house area. By adjusting $p$, we can generate several strategies, such as peer-to-peer strategy when $p$ is small or zone defense strategy when $p$ is large. We can also observe the intermediate strategies by tuning parameter $p$ to be between the two strategies.

\section{DISCUSSION}
In this section, we discuss the problems that can be further improved for this paper.

\begin{enumerate}
  \item We need to test the simulation with more scenes. In this paper, we have only tested the system with two general scenes. It is necessary to investigate the new logic further with several more scenes to confirm our assumption.
  \item We need to automate the extraction of data from the videos. In this paper, we extract the input manually from the video. The method can be inaccurate as a human estimate the data. Developing a technique to automate this process is necessary to gain accurate input for testing the logic.
  \item We need to take into consideration of the puck position in the future simulation. In the current simulation, we do not take into account the puck position. Although we assign the offensive player who holds the puck as a puck holder, it is undeniable that there exists a time when the puck is not with any players, such as when the puck holder sends the puck to the other player. Thus, the generated movement might not accurately reflect the actual game movement.
  \item We need to create an evaluation metric to standardize our evaluation. In this experiment, we asked the professional to evaluate our experiment's result. With ample experience in the field, there is some credibility to his evaluation. However, it is undeniable that there is a need to create a systematic evaluation metric for evaluating defensive movements.
\end{enumerate}

\end{document}